\documentclass{aa}
\usepackage{graphics,latexsym}
\begin{document}

   \thesaurus{02     
              (12.03.4;  
               12.04.1;  
               12.07.1;  
               12.12.1;  
	       03.13.4   
               )} 
   \title{Cosmic shear and halo abundances: analytical versus numerical results}

   \author{Katrin Reblinsky\inst{1} \and
           Guido Kruse\inst{1} \and
	   Bhuvnesh Jain\inst{2} \and
           Peter Schneider\inst{1}
          }

   \offprints{Katrin Reblinsky (reblinsk@mpa-garching.mpg.de)}

   \institute{Max-Planck Institut f\"ur Astrophysik, Postfach 1523,
              D-85740 Garching, Germany
         \and
             Johns Hopkins University, Department of Physics,
	     Baltimore, MD 21218, USA
             }

   \date{Received ; accepted }

   \authorrunning{K. Reblinsky et al.}

   \maketitle

\spaceskip=3.33pt plus 5.4pt minus 1.11pt

   \begin{abstract}

The aperture mass has been shown in a series of recent publications to
be a useful quantitative tool for weak lensing studies, ranging from
cosmic shear to the detection of a mass-selected sample of dark matter
haloes. Quantitative analytical predictions for the aperture mass have
been based on a number of simplifying assumptions. In this paper, we
test the reliability of these assumptions and the quality of the
analytic approximations, using ray-tracing simulations through a
cosmological density field generated by very large N-body
simulations. We find that those analytic predictions which take into
account the non-linear evolution of the matter distribution, such as
the dispersion of the aperture mass and the halo abundance, are
surprisingly accurately reproduced with our numerical results, whereas
the predictions for the skewness, based on quasi-linear theory, are
rather imprecise. In particular, we verify numerically that the
probability distribution of the aperture mass decreases exponentially
for values much larger than the rms. Given the good overall agreement,
comparisons between the observed distribution of the aperture mass and
the theoretical values provide a powerful tool for testing
cosmological models.

   \end{abstract}

%

\def\d{{\rm d}}

\section{Introduction}	

The gravitational distortion of light bundles from distant sources in
the universe provides a unique means to investigate (the statistical
properties of) the intervening mass distribution. Being observable
through the image distortion of the distant faint blue galaxy
population, this cosmic shear effect offers the opportunity to study
statistical properties of the large-scale structure. In contrast to
almost all other methods for investigating the large scale
structure (LSS) -- with CMB being the only exception -- no assumption
about the relation between dark and luminous mass is required.

Based on the assumption that the intrinsic orientation of the
background galaxies is random, a net alignment in the observed
galaxy images can be attributed to the tidal field (shear). Hence, the
alignment pattern of the galaxy images directly reflects the
properties of the mass distribution. For example, two-point
statistical measures of the galaxy ellipticities (understood here and
in the following as two-component quantities, with an amplitude and an
orientation) can be expressed directly in terms of the power spectrum
of the mass distribution, convolved with a filter function (Blandford
et al.\ 1991; Miralda-Escud\'e 1991; Kaiser 1992; Jain \& Seljak 1997;
Bernardeau et al.\ 1997; Schneider et al.\ 1998, hereafter SvWJK;
Kaiser 1998; van Waerbeke et al.\ 1999; Jain et al.\ 1999, hereafter
JSW; and references therein). The power spectrum completely
characterises a Gaussian random density field, and so the two-point
statistics, like the two-point correlation function of galaxy
ellipticities or the rms shear in an aperture, suffices to extract the
statistical information contained in the distorted galaxy
images. Whereas the earlier of the aforementioned papers concentrated
mainly on predictions for the cosmic shear based on the linear
evolution of the cosmic density field, it was pointed out by Jain \&
Seljak (1997) that even on scales as large as one degree, the
non-linear evolution significantly affects the expected amplitude of
the cosmic shear.

The non-linear evolution transforms an initially Gaussian field into a
non-Gaussian one, and thus the cosmic shear on small angular scales is
expected to display significant non-Gaussian features. As pointed out
by Bernardeau et al.\ (1997) and SvWJK, the skewness of the resulting
cosmic shear field is a sensitive measure of the density factor
$\Omega_0$, since in quasi-linear perturbation theory the skewness is
independent of the normalisation of the initial power spectrum.  In
order to define the skewness, mass reconstruction algorithms such as
those developed for cluster reconstructions (Kaiser \& Squires 1993;
Seitz \& Schneider 1996, Seitz et al.\ 1998a; Lombardi \& Bertin 1998a,b;
and references therein) can be employed to reconstruct the projected
density field. This density field, appropriately
spatially filtered, can then be used to calculate the skewness. In
contrast to the two-point statistical measures mentioned above, which
are defined directly in terms of the observable image ellipticities,
this measurement of skewness is more indirect. This causes 
estimates of the statistical error from the data itself not to be 
straightforward. On the other hand, the aperture mass $M_{\rm ap}$,
introduced as a measure for cosmic shear in SvWJK, is a scalar
quantity directly defined in terms of the image ellipticities, and can
thus be easily used for defining a skewness, as well as a second order
statistics, the rms of $M_{\rm ap}(\theta)$, where $\theta$ is the
angular scale of the circular aperture (definitions are given in
Sect.\ 2 below). In particular, in contrast to the two-point
ellipticity correlation function and the rms shear in an aperture,
for which the filter with which the power spectrum of the projected
density field is measured is broad, the corresponding filter
function for the rms of $M_{\rm ap}(\theta)$ is very narrow, and can be
approximated very accurately by a delta ``function'', so that $M_{\rm
ap}(\theta)$ directly measures the power of the projected density at
wavelength $\ell\approx 4.25/\theta$ (Bartelmann \& Schneider 1999).
The skewness defined in terms of $M_{\rm ap}$ has been considered in
SvWJK. 

As in the evolution of the three-dimensional density field, where
highly non-linear structures like clusters of galaxies form, the
projected mass density attains strongly non-Gaussian features, e.g.,
the projection of collapsed haloes. As a scalar quantity, the aperture
mass is ideally suited to probe the full probability distribution of
the projected mass density; in particular, values of $M_{\rm
ap}(\theta)$ far out in the non-Gaussian tail signal the presence of
massive dark matter haloes. Therefore, peaks in the distribution of
$M_{\rm ap}$ can be used to search for such haloes, independent of
their luminous properties (Schneider 1996, hereafter S96). Indeed, a
first application of $M_{\rm ap}$ to large scale structure
simulations by Reblinsky \& Bartelmann (1999) revealed that the
detection of dark matter haloes through the aperture mass is more
reliable in terms of completeness and spurious detections and suffers
less from projection effects than optically selected cluster
samples. Assuming that highly significant peaks are caused by such
haloes, one can predict their abundance (i.e., number of peaks above a
certain threshold per unit solid angle) by combining the spatial
abundance as predicted by Press \& Schechter (1974) theory with an
assumed density profile, such as the universal dark matter profile
found by Navarro et al.\ (1996, 1997; combined NFW). This idea has
been put forward by Kruse \& Schneider (1999a; hereafter KS1), who
found that, depending on the cosmological model and the redshift
distribution of background galaxies, of order 10 such haloes per
square degree will be detectable in deep ground-based optical images,
with a signal-to-noise ratio larger than 5. Given the rapid evolution
of wide-field imaging, a mass-selected sample of dark matter haloes is
now well within reach. Indeed, a first example of a shear-detected
mass concentration has recently been found by Erben et al. (1999).

All of these predictions on cosmic shear are made using simplifying
assumptions in order to make analytic progress. JSW tested some of
these assumptions using ray-tracing simulation through a cosmic
density field generated by very large N-body simulations; similar
tests have been carried out by van Waerbeke et al.\ (1999). They found
that the major approximations made in these analytic treatments,
namely the so-called ``Born-approximation'' (which projects the density
fields along ``straight lines''), and the neglect of non-linear terms in
the propagation equations, are very well satisfied. In particular, the
twist of light bundles, which vanishes identically in the usual
analytical treatments, is indeed very small. The predictions on the
projected power spectrum using the approximation for the fully
non-linear power spectrum of Peacock \& Dodds (1996) agree very well
with the numerical results, whereas the predictions concerning the
skewness are less precise, indicating a breakdown of quasi-linear
perturbation theory on small scales.

In this paper, we extend the study of JSW to the particular application
of the aperture mass statistics. We use the same simulations as 
JSW, resulting in a table of shear and projected mass density as a
function of angular position. From the shear, we can simulate
observations of the aperture mass as a function of position on the
``data'' field, and investigate its statistical properties. In
particular, we calculate the probability distribution of $M_{\rm
ap}(\theta)$, which we find to be highly non-Gaussian, and from that we
study the dispersion, skewness and kurtosis of the distribution. In
agreement with JSW, we find that the dispersion is accurately
predicted by analytic theory, whereas the skewness predictions can
differ substantially from the numerical results. Of particular
interest is the kurtosis, since it enters the determination of the
uncertainty of the dispersion measurement due to cosmic variance
(SvWJK). We find that the kurtosis is a slowly decreasing function of
angular scale $\theta$, and attains values of $\sim 3$ even on scales
as large as $10'$; hence, the cosmic variance will be the major source
of statistical error in the measurement of the power spectrum of the
projected density field from $M_{\rm ap}$. The analytic predictions
on the abundance of significant peaks of $M_{\rm ap}$, and thus
presumably of dark matter haloes, turns out to be remarkably precise,
given the strong assumptions made. We confirm the high
number density of haloes detectable with this method. The shape of the
probability distribution of $M_{\rm ap}$ in the highly non-Gaussian
tail, predicted by Kruse \& Schneider (1999b; hereafter KS2), can also
be confirmed to be well approximated by an exponential function.

\section{The aperture mass measure $M_{\rm ap}$}
\label{aperturemass}
In this section, we briefly summarise the properties of the aperture
mass, i.e., its definition, its relation to the shear, and its
signal-to-noise ratio. For more details, the reader is referred to S96
and SvWJK.

\subsection{$M_{\rm ap}$ statistics}
We define the spatially filtered mass inside a circular aperture of
angular radius $\theta$ around the point $\mbox{\boldmath$\zeta$}$ by
\begin{equation}
M_{\rm ap} (\mbox{\boldmath$\zeta$}):=\int \d^2 \vartheta 
\ \kappa(\mbox{\boldmath$\vartheta$})
\ U(\vert \mbox{\boldmath$\vartheta-\zeta$} \vert),
\label{mapt}
\end{equation}
where the continuous weight function $U(\vartheta)$ vanishes for
$\vartheta>\theta$.  If $U(\vartheta)$ is a compensated filter
function,
\begin{equation}
\int_0^{\theta} \d \vartheta \ \vartheta \ U(\vartheta)=0,
\end{equation} 
one can express $M_{\rm ap}$ in terms of the tangential shear
$\gamma_{\rm t}(\mbox{\boldmath$\xi$}; \mbox{\boldmath$\zeta$})$ at
position $\mbox{\boldmath$\xi + \zeta$}$
relative to $\mbox{\boldmath$\zeta$}$ as
\begin{equation}
M_{\rm ap} (\mbox{\boldmath$\zeta$})=\int \d^{2} \xi \ 
\gamma_{\rm t} (\mbox{\boldmath$\xi$}; \mbox{\boldmath$ \zeta$}) 
\ Q(\vert \mbox{\boldmath$\xi$} \vert),
\label{mapshear}
\end{equation}
(Fahlmann et al. \ 1994; S96), where
\begin{equation}
\gamma_{\rm t}(\mbox{\boldmath$\xi$}; \mbox{\boldmath$\zeta$}) 
= -{\rm Re}(\gamma(\mbox{\boldmath$\xi+\zeta$}) \, {\rm e}^{-2{\rm i} \phi}),
\label{tanshear}
\end{equation} 
and $\phi$ is the polar angle of $ \mbox{\boldmath$\xi$}$.
The function $Q$ is related to $U$ by
\begin{equation}
Q(\vartheta) = \frac{2}{\vartheta^2} \ \int_0^{\vartheta}
\d \vartheta^{\prime} \ \vartheta^{\prime} \ U(\vartheta^
{\prime})
\ - U(\vartheta) .
\end{equation}

We use the filter function for $l=1$
from the family given in SvWJK: writing 
$U(\vartheta)=u(\vartheta/\theta)/\vartheta^2$,
and $Q(\vartheta)=q(\vartheta/\theta)/\vartheta^2$, we take
\begin{equation}
\label{SvWJK98}
u(x)=\frac{9}{\pi} \left(1 - x^{2} \right) 
\left( \frac{1}{3} - x^{2} \right),
\end{equation}
and
\begin{equation}
q(x)=\frac{6}{\pi}x^2(1-x^2),
\end{equation}
with $u(x)=0=q(x)$ for $x>1$.

\subsection{Signal-to-noise ratio}
\label{sn_section}
 
An estimate of the shear field $\gamma$, and thus of the aperture
mass $M_{\rm ap}(\mbox{\boldmath$\vartheta$})$ through Eq.
(\ref{mapshear}), is provided by the distortions of images of faint
background galaxies. The complex ellipticity of galaxy images is
defined in terms of second moments of the surface-brightness tensor
(e.g., Tyson et al.\ 1990; Kaiser \& Squires 1993). Specifically, we
use here the ellipticity parameter $\epsilon$ (Schneider 1995;
Seitz \& Schneider 1997), which is defined such that for sources with
elliptical isophotes of axis ratio $r\le1$, the modulus of the
source ellipticities is given as $|\epsilon^{({\rm s})}|=(1-r)/(1+r)$,
and the phase of the $\epsilon^{({\rm s})}$ is twice the position
angle of the major axis.

The complex image ellipticity $\epsilon$ can then be calculated in terms
of the source ellipticity $\epsilon^{({\rm s})}$ and the reduced shear
$g\equiv\gamma\,(1-\kappa)^{-1}$ by the transformation (Seitz \&
Schneider 1997)
\begin{equation}
\epsilon=\frac{\epsilon^{({\rm s})} + g}{1+g^{*} \epsilon^{({\rm
s})}}\;.
\label{eps}
\end{equation}
This relation is valid only for noncritical clusters. For
critical clusters, it has to be replaced by a different
transformation. However, as we are mainly interested in the weak
lensing regime, the above relation is sufficient here.

It has been demonstrated (Schramm \& Kayser 1995; Seitz \& Schneider
1997) that the ellipticity $\epsilon$ of a galaxy image is an unbiased
estimate of the local reduced shear, provided that the intrinsic
orientations of the sources are random.  In the case of weak lensing,
$\kappa\ll1$, one then has
\begin{equation}
\langle \epsilon \rangle =g \approx \gamma 
\end{equation}
by averaging (\ref{eps}) with the probability distribution of the
source ellipticities.  

As for the tangential shear component $\gamma_{\rm t}$ occurring
in (\ref{tanshear}), a similar quantity for the image ellipticities can
be defined. Consider a galaxy image $i$ at a position 
$\mbox{\boldmath$\vartheta$}_i+\mbox{\boldmath$\zeta$}$ relative to
the point $\mbox{\boldmath$\zeta$}$ with a complex image ellipticity 
$\epsilon_i$. In analogy to (\ref{tanshear}) the tangential ellipticity
$\epsilon_{{\rm t}i}(\mbox{\boldmath$\vartheta_i$};
\mbox{\boldmath$\zeta$})$ of this galaxy is then given by 
\begin{equation}
\epsilon_{{\rm
t}i}(\mbox{\boldmath$\vartheta$}_i;\mbox{\boldmath$\zeta$})=-{\rm 
Re} \left( \epsilon_i
(\mbox{\boldmath$\vartheta$}_i+\mbox{\boldmath$\zeta$})
\ {\rm e}^{-2 {\rm i} \phi_i} \right),
\label{epsi_t}
\end{equation}
where $\phi_i$ is the polar angle of $\mbox{\boldmath$\vartheta$}_i$.

We can now estimate the integral (\ref{mapshear}) by a
discrete sum over galaxy images,
\begin{equation}
M_{\rm
ap}(\mbox{\boldmath$\zeta$})=\frac{1}{n}\,\sum_{i}\epsilon_{{\rm
t}i}(\mbox{\boldmath$\vartheta$}_i;\mbox{\boldmath$\zeta$})\, 
Q(|\mbox{\boldmath$\vartheta$}_i|),
\label{map_1}
\end{equation}
where $n$ is the number density of galaxy images.
The discrete dispersion $\sigma_{\rm d}$
of the aperture mass 
$M_{\rm ap}(\mbox{\boldmath$\zeta$})$ is found by squaring
(\ref{map_1}) and taking the expectation value in the absence of
lensing, which leads to  
\begin{equation}
\label{disp_d}
\sigma_{\rm
d}^{2} = \frac{\sigma_\epsilon^2}{2\,n^2}  
\sum_{i}Q^2(|\mbox{\boldmath$\vartheta$}_i|),
\end{equation}
where $\sigma_\epsilon^2=\langle \vert \epsilon^{({\rm s})} \vert^2 \rangle$.
Performing an ensemble average of Eq. (\ref{disp_d}) leads to the continuous
dispersion $\sigma_{\rm c}$
\begin{equation}
\label{disp_c}
\sigma_{\rm c}^{2} (\theta) = \frac{\pi \sigma_{\epsilon}^2}{n} 
\int_{0}^{\theta} {\rm d} \vartheta \ \vartheta \ Q^{2}(\vartheta ).
\end{equation} 
Finally, the {\em signal-to-noise\/} ratio $S$ at position
$\mbox{\boldmath$\zeta$}$ is
\begin{equation}
S(\mbox{\boldmath$\zeta$}) \equiv \frac{M_{\rm
ap}(\mbox{\boldmath$\zeta$})}{\sigma_{\rm d}} = 
\frac{\sqrt{2}}{\sigma_\epsilon} \
\frac{\sum_i\epsilon_{{\rm t}i}(\mbox{\boldmath$\vartheta$}_i
;\mbox{\boldmath$\zeta$}) \
Q(|\mbox{\boldmath$\vartheta$}_i|)}
{\left[\sum_iQ^2(|\mbox{\boldmath$\vartheta$}_i|)\right]^{1/2}}\;.
\label{eq_sn} 
\end{equation}

In the simulations, we draw the source ellipticities from a Gaussian
probability distribution,
\begin{equation}
\label{distrib}
p_{\rm s}(|\epsilon^{({\rm s})}|)=
  \frac{1}{\pi\sigma_{\epsilon}^2
           \left[1-\exp\left(-\sigma_\epsilon^{-2}\right)\right]}\,
  \exp\left(-\frac{|\epsilon^{({\rm s})}|^2}{\sigma_\epsilon^2}\right)\;,
\end{equation}
where the width of the distribution is chosen as
$\sigma_\epsilon=0.2$. Throughout this paper, we assume that the
number density of the background sources is $n=30\,{\rm arcmin}^{-2}$.

\subsection{Analytical work done with $M_{\rm ap}$}
\label{map_anal}

The aperture mass has been considered in the framework of blank field
surveys in a variety of earlier publications. Introduced as a
convenient statistics for cosmic shear, SvWJK have calculated the rms
of $M_{\rm ap}$ as a function of angular scale, using the Peacock \&
Dodds (1996) approximation for the non-linear evolution of the power
spectrum of density fluctuations.
Like other two-point
statistics, the dispersion of $M_{\rm ap}$ is an integral over the
power spectrum of the projected mass distribution, weighted by a
filter function. The filter function corresponding to $\langle M_{\rm
ap}^2\rangle$ is very narrow and can be well approximated by a delta
function (Bartelmann \& Schneider 1999). Hence, $\langle M_{\rm
ap}^2(\theta)\rangle$ reproduces the shape of the projected power
spectrum and, depending on the cosmological model and the redshift
distribution of the sources, it reveals a broad peak at $\theta\sim
1'$. One convenient property of the aperture mass is that the correlation
function of $M_{\rm ap}$ of two apertures spatially separated by
$\Delta\theta$ quickly decreases and already achieves values of
$10^{-2}$ for $\Delta\theta\sim 2\theta$. This means that 
measurements of $M_{\rm ap}$ from a large consecutive area can be
considered independent if the apertures are densely laid out on this data
field; this is in contrast to the rms shear in apertures which is
strongly correlated, and thus must be obtained from widely separated
regions on the sky.

Being a scalar quantity, $M_{\rm ap}$ can also be used for
higher-order statistical measures of the cosmic shear. SvWJK
calculated the skewness of $M_{\rm ap}$, using Eulerian perturbation
theory for the evolution of the three-dimensional density contrast
$\delta$. In agreement with Bernardeau et al.\ (1997) they found that
the skewness is a sensitive function of the cosmic density factor
$\Omega_0$, and is in this approximation independent of the
normalisation of the power spectrum. 

A measurement of the dispersion of $M_{\rm ap}$ is affected by two
main sources of statistical error: the intrinsic ellipticity
distribution of the source galaxies, and cosmic variance. To estimate
the latter, one needs to know the kurtosis of $M_{\rm ap}$ which
cannot easily be determined analytically.

Values of $M_{\rm ap}$ much larger than its rms probe the highly
non-Gaussian regime of the projected density field. From its
definition, one sees that large values of $M_{\rm ap}$ are expected if
the aperture is centred on a density peak with a size comparable to
the filter scale $\theta$. Therefore, the aperture map can be used to
search for such density peaks, presumably collapsed dark matter
haloes, in blank field imaging surveys. In this way it is possible to obtain a
mass-selected sample of such haloes (S96). Simple analytical arguments
in S96 suggest that dark matter haloes with an approximately
isothermal profile are detectable with a signal-to-noise ratio
larger than 5 if their velocity dispersion exceeds $\sim
600$ km/s, assuming a number density of background sources of $n\sim
30$ arcmin$^{-2}$. Indeed, this theoretical expectation was verified
in the lensing investigation of the cluster MS1512+36 (Seitz et al.\
1998b). This cluster has a velocity dispersion of about $\sim 600$
km/s, as obtained from strong lensing modelling and from spectroscopy
of cluster members, and is detected in the weak lensing analysis with
very high statistical significance.

Assuming that the high signal-to-noise peaks of $M_{\rm ap}$ are due
to collapsed dark matter haloes, one can attempt to estimate the
abundance of such peaks using analytic theory. KS1 have calculated the
number density of haloes with aperture mass larger than $M_{\rm ap}$,
$N(> M_{\rm ap}, \theta)$, assuming (1) that dark matter haloes are
distributed according to Press \& Schechter (1974) theory which
yields the number density of collapsed haloes as a function of halo
mass and redshift, and (2) that the azimuthally-averaged projected
density profiles of these haloes can be described by the projection of
the universal halo density profile found in numerical simulations by
NFW. Depending on the cosmological model and on the redshift distribution
of the faint galaxies, the number density of peaks of $M_{\rm ap}$
with a signal-to-noise ratio larger than 5 was estimated to be $\ga
10$ per square degree, and the redshift distribution of these haloes
is strongly dependent on the behaviour of the linear growth factor for
density perturbations, and thus on $\Omega_0$. This abundance is
encouraging, since it allows one to obtain samples of haloes selected
by their mass properties alone (for a first example, see Erben et al.\
1999).

Using the same model, KS2 have calculated the probability distribution
of $M_{\rm ap}$ for values of $M_{\rm ap}$ much larger than its rms,
assuming that this non-Gaussian tail of the probability distribution
is dominated by dark matter haloes. They found that the distribution
is very well described by an exponential; i.e., the tail is much broader
than for a Gaussian.

All these analytic predictions are based on a number of
approximations and simplifying assumptions. In Sect.\ 4 below we
shall compare these analytic results with those found in
ray-tracing simulations through a cosmological mass distribution
obtained from very large N-body calculations, as described in the next
section.

\section{Generation of shear maps with ray-tracing simulations}
\label{ray_tracing}

Simulated shear maps due to weak lensing by large-scale structure
are made by performing ray tracing simulations through the dark
matter distribution produced by N-body simulations (JSW). 
The N-body simulations used  are a set of adaptive
particle-particle/particle-mesh (AP$^3$M) simulations. 
The long-range component of the gravitational force is computed 
by solving Poisson's equation on a grid. The grid calculation is supplemented 
with a short range correction computed either by a direct sum over
neighbouring particles, or, in highly clustered regions, by combining a
calculation on a localised refinement mesh with a direct sum over
a smaller number of much closer neighbours. 
The parameters used by the N-body simulations are given in Table \ref{cosmo}.

The simulations were run with a parallel adaptive AP$^3$M code
(Couchman et al. 1995; Pearce \& Couchman 1997) kindly made
available by the Virgo Supercomputing Consortium (e.g. Jenkins et al.
1998). They followed $256^3$ particles using a force law with
softening length $l_{\rm soft}\simeq 30\ h^{-1}$kpc at $z=0$ (the
force is $\sim 1/2$ its $1/r^2$ value at one softening length and is
almost exactly Newtonian beyond two softening lengths). $l_{\rm soft}$
was kept constant in physical coordinates over the redshift range of
interest to us here. The simulations were carried out using 128 or 256
processors on CRAY T3D machines at the Edinburgh Parallel Computer
Centre and at the Garching Computer Centre of the Max-Planck
Society. These simulations have previously been used for studies of
strong lensing by Bartelmann et al. (1998), for studies of dark matter
clustering by Jenkins et al. (1998), and for studies of the relation
between galaxy formation and galaxy clustering by Kauffmann et al.
(1999a,b), and Diaferio et al. (1999).

The ray tracing simulations of weak lensing from which we use the convergence
and shear maps were computed by JSW. They used a 
multiple lens-plane calculation that implements the discrete recursion
relations for the position of a given photon and for the Jacobian matrix 
of the lens mapping at this position (Schneider \& Weiss 1988; 
Schneider et al. 1992; see Seitz et al.\ 1994 for a thorough
justification for this approach). Aside from the distance factors, 
the main input into the recursion relations is the shear matrix at
each lens plane. 
The ray tracing algorithm consists of three parts: constructing
the dark matter lens planes, computing the
shear matrix on each plane, and using these to evolve the photon
trajectory from the observer to the source. The details involved
at each step are as follows: 

\noindent 1. The dark matter distribution between source and
observer is projected onto $20-30$ equally spaced (in comoving
distance) lens planes.  The particle positions on each plane are
interpolated onto a grid of size $2048^2$. Since the three-dimensional
mass distribution is taken from a single realisation of the evolution
of the LSS, the projected mass distributions of consecutive lens
planes are correlated. In order to decorrelate them, the
projection is carried out along a randomly chosen one of the three
coordinate axes; in addition, the origin of the coordinate system
in each lens plane is translated by a random vector and the lens plane
is rotated by a random angle. In this way, the projected mass
distributions of consecutive lens planes are as independent as
possible, given the restriction of only a single realisation of the
3-d matter distribution.

\noindent 2. On each plane, the shear matrix is computed on a grid by 
Fourier transforming the projected density
and using its Fourier space relation to the shear. The inverse
Fourier transform is then used to return to real space. 

\noindent 3. The photons are started on a regular grid on the first
lens plane. Perturbations along the line of sight distort this grid
and are computed using the relation between deflection angle and
projected density. Once we have the photon positions, we
interpolate the shear matrix onto them and solve the recursion relations
for the Jacobian of the mapping from the $n$-th lens plane
to the first plane. 

\noindent 4. Solving the recursion relations up to the source plane
yields the Jacobian matrix at these positions.  Note that the ray
tracing is done backwards from the observer to the source, thus
ensuring that all the photons reach the observer. The first lens plane
is the image plane and has the unperturbed photon positions. All
sources are assumed to be at a redshift of $z_{\rm s}=1$.

There are two kinds of resolution limitations in the ray-tracing simulations. 
The first reflects the finite size and resolution of our N-body simulations,
the second the use of finite grids when computing deflection angles
and shear tensors on the lens planes. 
At the peak redshift of the lensing contribution, both effects give a
small scale resolution of order $0.2'$. However, since the 
lens efficiency is not very sharply peaked, effects at other 
redshifts also enter. Thus depending on the statistical measure being
used, the small scale resolution lies in the range $\sim 0.2'-0.4'$. 

On large scales the finite box-size of the N-body simulations sets
the upper limit on the angular scales available. 
The angular size of our simulation box at $z=1$ is about 3$^\circ$. 
Thus on scales comparable to $1^\circ$, only a few sample regions are 
available, leading to large fluctuations across different
realisations. We therefore restrict our considerations to apertures
with radius $\theta\le 10'$ using one realisation for every 
cosmological model. For the $\tau$CDM model, we use ten
different realisations of the ray tracing simulations (i.e., they
differ in the direction of  projections, the translation and rotation
of the projected matter distribution in the individual lens planes) to 
estimate the cosmic variance.
 
\begin{table}
\centering
\caption[]{\label{cosmo} 
Parameters of the N--body simulations.}
\begin{tabular}{@{}rllll@{}}
\hline \\
  \multicolumn{1}{c}{Simulation} &
  \multicolumn{1}{c}{SCDM} &
  \multicolumn{1}{c}{$\tau$CDM} &
  \multicolumn{1}{c}{$\Lambda$CDM} &
  \multicolumn{1}{c}{OCDM} \\ [10pt]\hline \\
  $N_{\rm par}$                 & $256^3$ & $256^3$ & $256^3$ & $256^3$ \\
  $l_{\rm soft} [h^{-1}$ kpc]   & 36 & 36 & 30 & 30 \\
  $\Gamma$                      & 0.5 & 0.21 & 0.21 & 0.21 \\
  $L_{\rm box}  [h^{-1}$ Mpc]   & 85 & 85 & 141 & 141 \\
  $\Omega_0$                    & 1.0 & 1.0 & 0.3 & 0.3 \\
  $\Lambda_0$                   & 0.0 & 0.0 & 0.7 & 0.0 \\
  $H_0$ [km/s/Mpc]              & 50 & 50 & 70 & 70 \\
  $\sigma_8$                    & 0.6 & 0.6 & 0.9 & 0.85 \\
  $m_{\rm p} 10^{10} h^{-1}{\rm M}_{\sun}$ & 1.0 & 1.0 & 1.4 & 1.4 \\
  field size [$^{\circ}$]         & 2.7&2.7&3.4&3.9 \\
  \hline \\
\end{tabular}
\end{table}


\section{Application of $M_{\rm ap}$ to simulated shear maps}

For each of the shear maps generated as described in the last section,
we create a 2-dim. ``$M_{\rm ap}$ map'' by simulating ``observations''
of $M_{\rm ap}$ as a function of position on the 2-dim. shear
maps. The probability distribution function of $M_{\rm ap}$ (PDF) and
some of its moments are then calculated for every $M_{\rm ap}$ map and
compared to the analytical model.
\begin{figure}[!ht]
\resizebox{\hsize}{!}{\includegraphics{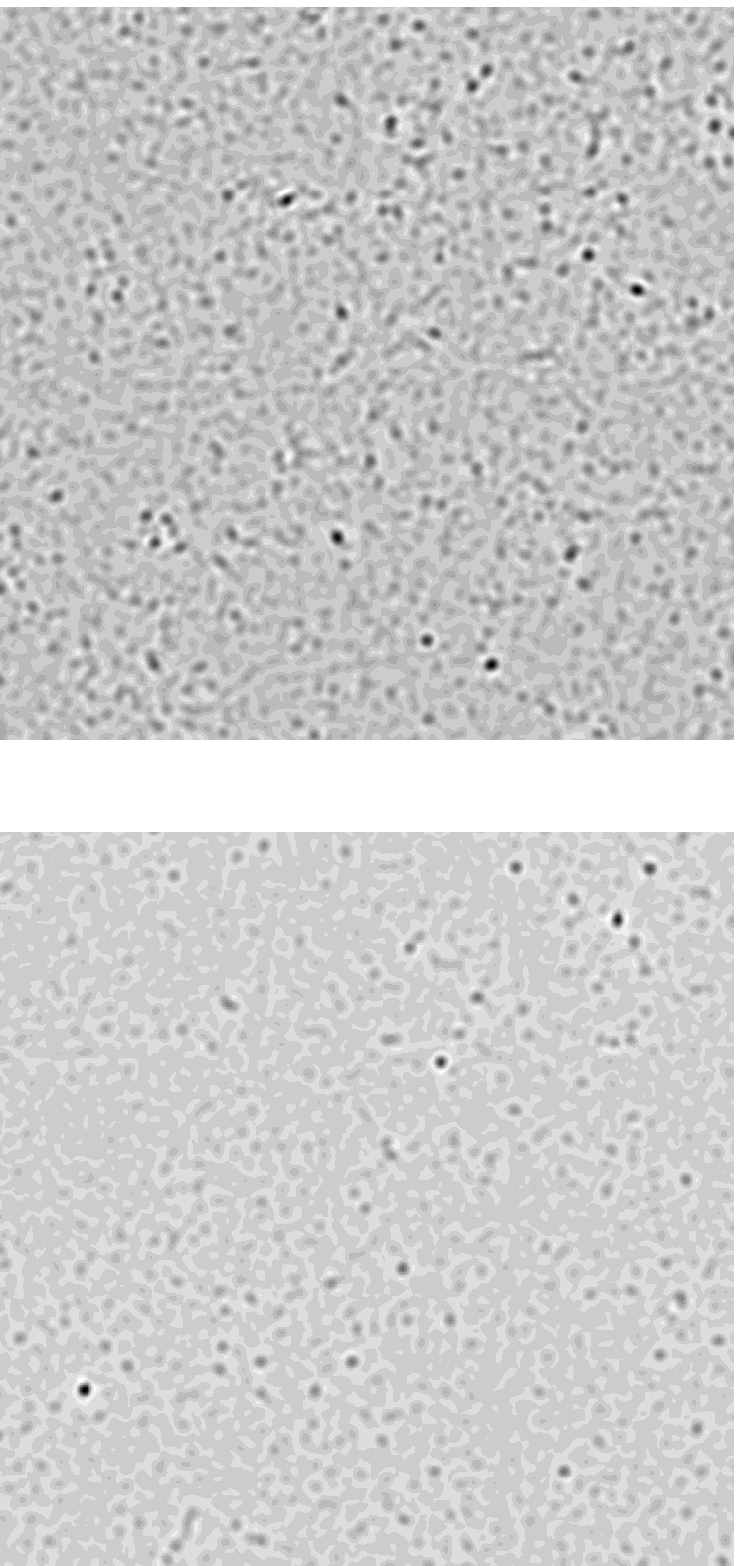}}
\caption{The 2-dimensional distribution of $M_{\rm ap}$ for a standard
CDM (SCDM, upper panel) and an open model (OCDM, $\Omega_{\rm m}=0.3$,
lower panel), with parameters given in Table 1.
The field size in both panels is 2$^{\circ}$.
\label{s_and_ocdm}}
\end{figure}

It is most instructive to consider two different sets of simulated
maps: in the first, we neglect noise from the
intrinsic ellipticity distribution of the background sources and
compute $M_{\rm ap}$ directly from the shear values on the grid
according to Eq. (\ref{mapshear}). We do this either in the limit of
weak lensing, i.e. we use (\ref{mapshear}) directly, or
we replace $\gamma_{\rm t}$ in (\ref{mapshear}) by the reduced shear
$g_{\rm t}$, which is the quantity estimated from the observable galaxy
ellipticities.

In the second set of simulations we introduce ellipticities of 
background galaxies according to the distribution function (\ref{distrib}).
The ellipticities add noise to $M_{\rm ap}$.

The noise-free results are the ones best compared to the analytic
results, whereas the ones accounting for intrinsic ellipticities yield
a more realistic description of the observational situation. In the
following the term ``without noise'' will refer to the first set of
$M_{\rm ap}$ simulations, while the term ``with noise'' will be used
for the second one.

As an illustrative example, the 2-dimensional distribution of $M_{\rm
ap}$ for a standard CDM (SCDM) and an open model (OCDM) is shown in
Fig. \ref{s_and_ocdm}. In both cases high peaks in these maps
correspond to haloes in the intervening matter distribution. It is possible 
to construct a shear-limited sample of haloes from these maps and 
to determine their abundance.

Comparing the two model universes, we see that the $M_{\rm ap}$ maps reflect 
the different growth of structure in different cosmologies. The 
$M_{\rm ap}$ map of the OCDM model is dominated
by many isolated peaks which correspond to already collapsed dark
matter haloes. The level of background noise coming from matter not yet
collapsed is considerably smaller than for the SCDM model in which the
structure forms later. The peaks in the SCDM model are less 
pronounced and isolated than in the open model.

\subsection{The PDF of $M_{\rm ap}$ and its moments}

Once we have computed the 2-dimensional distribution of $M_{\rm ap}$,
it is straightforward to determine the one-point probability
distribution function (PDF) of $M_{\rm ap}$ and its moments. The PDF
contains the cosmological information. The lower order moments like
rms value and skewness can be derived analytically under simplifying
assumptions, but the PDF itself cannot be calculated.  Therefore, ray
tracing simulations provide the only tool for testing the precision of
the analytical calculations.

\begin{figure*}[!ht]
\resizebox{12cm}{!}{\includegraphics{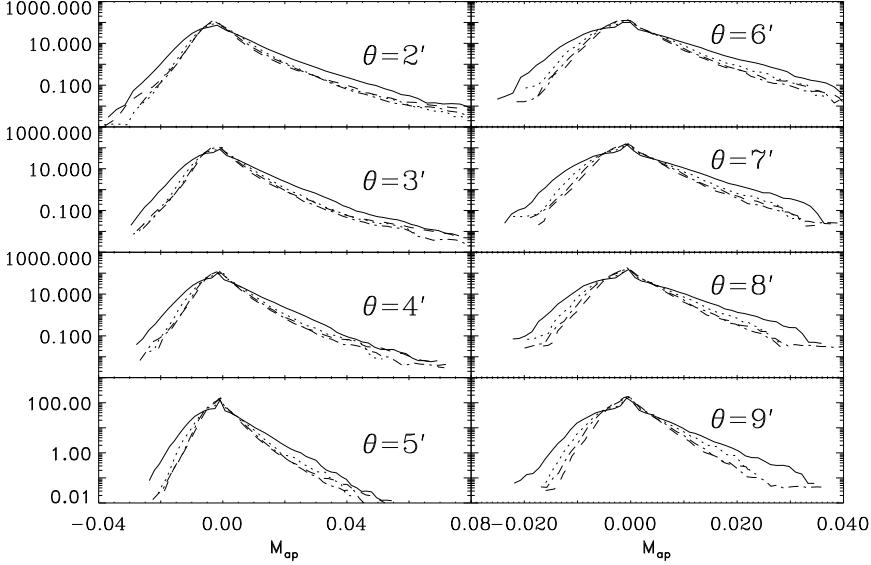}}
\hfill
\parbox[b]{55mm}{
\caption{The normalised PDF of $M_{\rm ap}$ for different filter
scales $\theta$ and cosmologies:
SCDM (solid line), $\tau$CDM (dotted line), OCDM (dashed line) and
$\Lambda$CDM (dashed-dotted line). The histograms are obtained from
$M_{\rm ap}$ maps without noise. Note the different scales on the
horizontal axis.
\label{histo}}}
\end{figure*}
\begin{figure*}[!ht]
\resizebox{12cm}{!}{\includegraphics{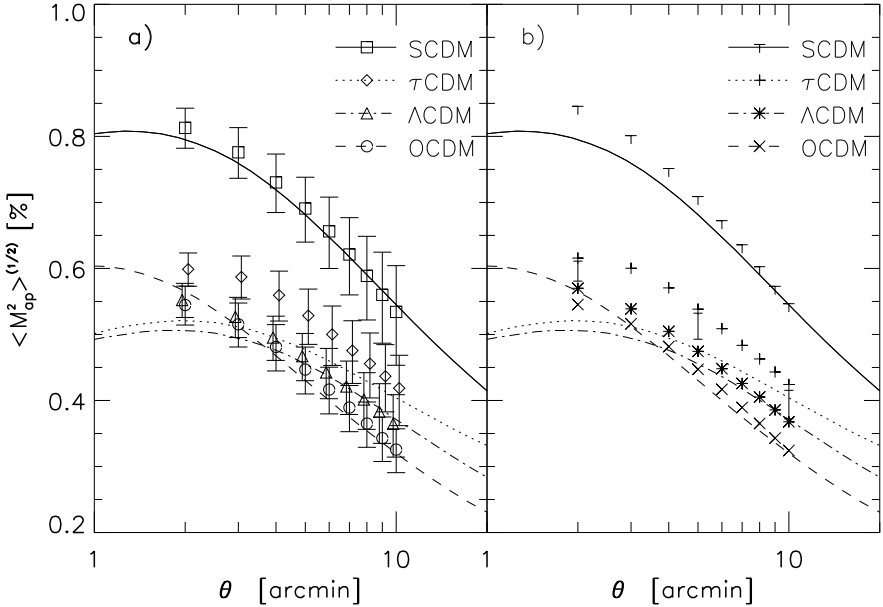}}
\hfill
\parbox[b]{55mm}
{\caption{ The rms value of $M_{\rm ap}$ computed with the filter
(\ref{SvWJK98}) versus filter scale $\theta$ for different
cosmologies. Lines refer to analytic values of $\langle M_{\rm
ap}^{2} \rangle^{1/2}$ from SvWJK, while symbols refer to rms values
obtained from simulations without noise, using $\gamma_{\rm t}$ (left panel)
and $g_{\rm t}$ (right panel). The error bars in the left panel are
determined from (\ref{sigma_map}) and (\ref{sigma_map_field}). The
symbols for OCDM and $\tau$CDM are slightly offset on the
$\theta$-axis for better display. The error bars in the right panel
show the standard errors from 10 realisations of the $\tau$CDM model
for 2,5, and 10 arcmin. They are centred on the arithmetic mean ({\em
not} on the realisation plotted).
\label{dispersion}}}
\end{figure*}

The qualitative features of the PDF for different filter scales
$\theta$ and for the four different cosmologies (Table \ref{cosmo}) can be
studied in Fig. \ref{histo}. The first point to note is that the
non-Gaussian features, namely the tail of the PDF at high $M_{\rm ap}$
values, are less pronounced for larger filter scales. This is due to
the fact that the smaller filter scales are more sensitive to the
already collapsed, non-linear objects.  The second feature to note is
the exponential decrease of the tail of $M_{\rm ap}$ which was already
obtained semi-analytically in KS2. We shall
discuss this feature in more detail later in this section.

We now turn to the rms value $\langle M_{\rm ap}^2 \rangle ^{1/2}$ of
$M_{\rm ap}$. Fig. \ref{dispersion} compares the analytical rms value
of $M_{\rm ap}$ calculated using the nonlinear power spectrum of
Peacock \& Dodds (1996) to the rms values computed from the PDFs
without noise for $\gamma_{\rm t}$ (left panel) and $g_{\rm t}$ (right
panel). The comparison of the latter shows that the difference between
shear and reduced shear is negligible even on filter scales as small
as $\theta\sim2$ arcmin corresponding to the highly nonlinear regime
of the mass distribution.

In the left panel of Fig. \ref{dispersion}, there is an excellent
agreement between the analytic predictions and the rms values
computed from simulations for the SCDM model.  There is also good
agreement for the $\Lambda$CDM and OCDM models, especially for the
larger apertures.  The notable exception is the $\tau$CDM model, for
which the simulations for small filters deviate by a larger factor
from the theoretical predictions.

When interpreting this difference between analytical calculation and
simulation in the $\tau$CDM model, one has to keep in mind that the
numerical results of Fig. \ref{dispersion} are based on a single
realisation. As the cosmic variance is relatively large, it is possible
that the large deviation is due to the special choice of the
realisation. This interpretation is supported by the fact that the
mean for the 10 realisations is considerable lower than for the single
realisation plotted. Furthermore, the field sizes of the simulated
fields used are too small to represent a characteristic region of the
universe.

\begin{figure}[!ht]
\resizebox{\hsize}{!}{\includegraphics{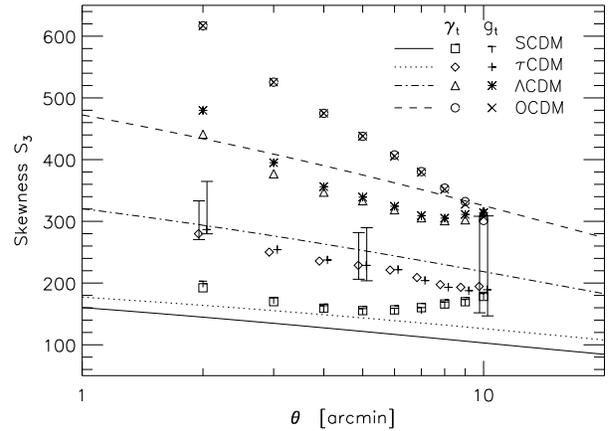}}
\caption{ The skewness $S_3$ of the PDF of $M_{\rm ap}$ as defined in
(\ref{skew_def}) as a function of filter scale $\theta$ for the same
cosmological models as in Fig. \ref{histo}. The analytical skewness
(lines) from quasi-linear theory is compared to the skewness obtained
from the PDF for both, the tangential shear $\gamma_{\rm t}$ and the reduced
shear $g_{\rm t}$. Errors on the $\tau$CDM model like in Fig.
\ref{dispersion}.
\label{skew}}
\end{figure}

The next higher moment of the PDF is the skewness, which is 
defined as 
\begin{equation}
\label{skew_def}
S_{3}(\theta):=\frac{\langle M_{\rm ap}^3\rangle}{\langle M_{\rm ap}^2\rangle^2},
\end{equation}
for which we can perform a similar analysis as for the rms value of
$M_{\rm ap}$. As pointed out by Bernardeau et al. (1997), van Waerbeke
et al. (1999), and JSW, the skewness defined in analogy to
(\ref{skew_def}) using a top-hat filter is a very sensitive probe of
the cosmic density parameter $\Omega_0$.

The dependence of the skewness on filter scale $\theta$
is displayed in Fig.  \ref{skew}. Again, we compare the skewness
computed from the PDF obtained from the ray tracing simulations
without noise, both using $\gamma_{\rm t}$ and $g_{\rm t}$, to the skewness of
$M_{\rm ap}$ obtained using quasi-linear theory (SvWJK).  The error
bars on the skewness for the $\tau$CDM model for 2, 5, and 10 arcmin
are derived from the 10 different realisations and are centred on
their arithmetic mean.

Again, the differences between the skewness obtained from simulations with
$\gamma_{\rm t}$ and $g_{\rm t}$ are small, though slightly larger than
for the dispersion, owing to the larger contribution from
high-$\kappa$ regions to the skewness. This difference, which is of
order a few percent at most, has been predicted to be small in the
Appendix of SvWJK.

When comparing the skewness as determined from second-order
perturbation theory for the density evolution to that obtained from
simulations (either computed with $\gamma_{\rm t}$ or $g_{\rm t}$) we
see that the former underpredicts the skewness by factors of up to
2. This failure of quasi-linear theory for the prediction of
higher-order moments has been demonstrated previously (Jain \& Seljak
1997; Gaztanaga \& Bernardeau 1998).  As we only determine the
skewness on scales below 10 arcmin, we are in a regime where the
density contrast is non-linear already. The skewness as calculated by
Hui (1999) using the so-called hyper-extended perturbation theory
(Scoccimarro \& Frieman 1999) may provide a more accurate analytical
prediction of $S_3$ than that from second-order perturbation theory.

Another point to note is the increase of the skewness towards smaller
filter scales. Generally speaking, such a behaviour is expected, as
the non-linear structure growth becomes more and more important for
small filter scales. This increase is described insufficiently by
quasi-linear theory: for the two EdS universes and even for the
$\Lambda$ model on large filter scales above 5 arcmin, this increase
(not the absolute value!)  is predicted satisfactorily, but the slope
for the open model is larger than analytic values on all scales
displayed. This discrepancy between fully non-linear simulations and
quasi-linear theory can be attributed to the fact that the open model
is much more dominated by already collapsed, non-linear objects than
all other models.

\begin{figure}[!ht]
\resizebox{\hsize}{!}{\includegraphics{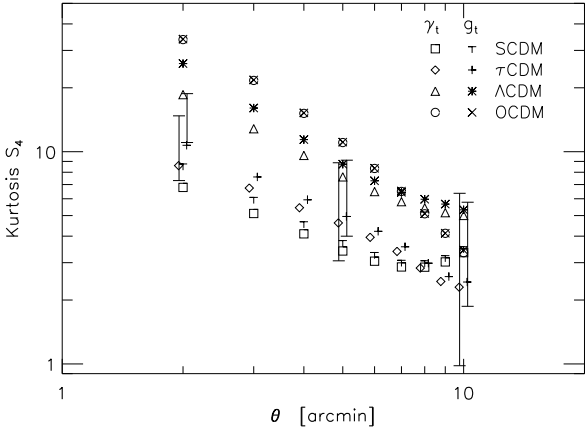}}
\caption{The kurtosis $S_4$ of the PDF of $M_{\rm ap}$ [Eq.\
(\ref{curt_def})] as a function of filter scale $\theta$, for the four
cosmological models. The kurtosis is derived from the PDF for both,
the tangential shear $\gamma_{\rm t}$ and the reduced shear $g_{\rm
t}$.  Errors on the $\tau$CDM model are like in
Fig. \ref{dispersion}. No analytic estimate of the kurtosis has been
calculated.
\label{curt}}
\end{figure}

The highest moment we consider explicitly is the kurtosis $S_4$
\begin{equation}
\label{curt_def}
S_{4}(\Theta):=
    \frac{\langle M_{\rm ap}^4\rangle}{\langle M_{\rm ap}^2\rangle^2}-3.
\end{equation}
The kurtosis is not only
important by itself, but also for
the determination of the error of the rms value of $M_{\rm ap}$, as
will be
discussed. As for the skewness, the kurtosis for the noise-free
simulations for both $\gamma_{\rm t}$ and $g_{\rm t}$ is plotted,
and the scatter for $\tau$CDM is determined from the 10 realisations.
No analytic result for $S_4$ is available; however, using third-order
perturbation theory, Bernardeau (1998) has calculated the kurtosis for
a top-hat filter.

We clearly see that the difference between $\gamma_{\rm t}$ and
$g_{\rm t}$ becomes important for the kurtosis, at least for the
smaller filter scales, since it is even more dominated by the
non-Gaussian tail of the PDF than the skewness.  The large error bars
on the kurtosis are mainly due to large cosmic variance in combination
with the small fields used; thus, the current simulations are unable
to provide an accurate determination of $S_4$.

We now turn to the error bars on the rms values of $M_{\rm ap}$ in
Fig. \ref{dispersion}. In the right panel, they were
estimated as the standard deviation from 10 different realisations for
the $\tau$CDM model. The error bars in the left panel were calculated
as follows:

As shown in SvWJK, an unbiased estimator of $\langle M_{\rm
ap}^2\rangle$ from a single aperture is
given by
\begin{equation}
M={(\pi\theta^2)^2\over N(N-1)}\sum_{i,j\ne i}^N Q_i\,Q_j\,
\epsilon_{{\rm t}i}\,\epsilon_{{\rm t}j} \;,
\end{equation}
where $N$ is the number of galaxies in the aperture, and $Q_i$ is the
value of the weight function $Q$ for the $i$-th galaxy. The dispersion
of this estimator is
\begin{equation}
\label{sigma_map_field}
\sigma^2(M) \approx S_4 \langle M_{\rm ap}^2\rangle^2
          + \left(\frac{6\sigma^2_{\epsilon}}{5\sqrt{2}N} 
	          +\sqrt{2}\langle M_{\rm ap}^2\rangle^2
	    \right)^2\;,
\end{equation}
where the two terms in parenthesis correspond to the noise from the
intrinsic ellipticity distribution, and the Gaussian cosmic variance,
respectively, whereas the term involving $S_4$ is the excess
cosmic variance due to non-Gaussianity. For a collection of $N_{\rm
f}$ independent apertures, all containing the same number of galaxy
images, an unbiased estimator for $\langle M_{\rm
ap}^2\rangle$ is the mean ${\cal M}$ of $M$ over these apertures, and the
dispersion is
\begin{equation}
\label{sigma_map}
\sigma\left({\cal M}\right) = \frac{\sigma (M)}{\sqrt{N_{\rm f}}}.
\end{equation}

Note that this result does not assume that the density field is
Gaussian. If one had a collection of $N_{\rm f}$ fields widely
separated on the sky, they would be statistically independent, so that
$N_{\rm f}=N$. In the opposite situation where a consecutive area on
the sky is available, one can lay down apertures on that field, but
they will not be statistically independent. However, as was shown in
SvWJK, the $M_{\rm ap}$ values of two apertures which touch each other
(i.e., with separation twice their radii), are almost
uncorrelated. Whereas the fact that the two aperture masses in these
two apertures are uncorrelated does {\it not} imply that they are {\it
independent} (which would mean that the joint probability distribution
for the values of $M_{\rm ap}$ would factorize) -- as would be the
case for Gaussian fields -- we assume the statistical independence for
estimating the effective number of fields $N_{\rm f}$ entering
(\ref{sigma_map}). Thus, the error bars in the left panel of
Fig. \ref{dispersion} are obtained from (\ref{sigma_map}), assuming
that the number of independent apertures is $N_{\rm f}=[\Theta/(2
\theta)]^2$, where $\Theta$ is the side length of the simulated shear
field.

In contrast, the error bars plotted in the right panel of Fig.
\ref{dispersion} for the $\tau$CDM model at the three different filter
scales $\theta=2,5,10$ arcmin are based on 10 different realisations
of the ray-tracing simulations and allow one to obtain a rough estimate for
the error from cosmic variance.  Notice that the error bars are
centred on the arithmetic mean of the 10 realisations and {\em
not} on the plotted results from a single realisation.

Comparing the size of the error bars in both panels, we see that both
methods give errors of the same order of magnitude even though the
errors estimated from the 10 realisations are smaller than the errors
from the estimator of $\langle M_{\rm ap}\rangle$. There are two
possible reasons for this: first, the effective number of independent
apertures is probably larger than our estimate given above, so that
the error bars on the left panel in Fig. \ref{dispersion} most likely
overestimate the true error. Second, in the calculation of the error
bars in the right panel, it was assumed that the 10 realisations are
independent; but as argued in Sect. \ref{ray_tracing} it is possible
that the realisations are not completely independent. This would lead 
to an underestimation of the cosmic variance. From Fig. \ref{dispersion}
these two competing effects cannot be quantified. It should be
noted that at least on the largest scale plotted, the contribution of
the intrinsic ellipticity distribution to the error
(\ref{sigma_map}) is completely negligible compared to the cosmic
variance.

\subsection{Halo abundances}

\begin{figure*}
\resizebox{12cm}{!}{\includegraphics{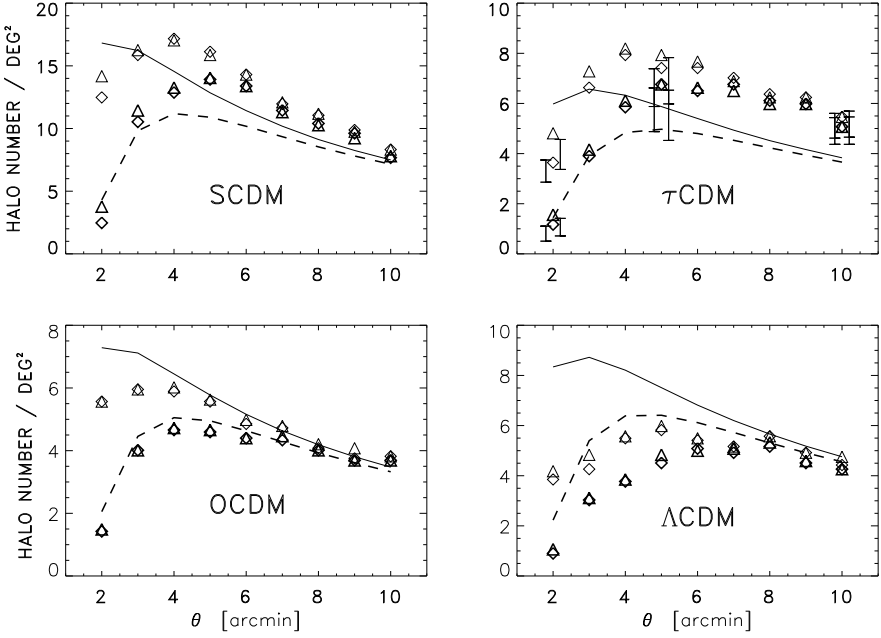}}
\hfill
\parbox[b]{55mm}{
\caption{The halo number density $N(> M_{\rm ap}, \theta)$ (thin
symbols and thin solid curves) and $N(> M_{\rm ap},> 0.6' ,\theta)$
(thick symbols and thick dashed curves) computed without noise as a
function of the filter scale $\theta$ for four cosmologies as
indicated in the panels. Symbols denote results from the simulations
($\Diamond$ from $\gamma_{\rm t}$, $\bigtriangleup$ from $g_{\rm t}$) whereas the
two line types display the corresponding analytic results from KS1. A
signal-to-noise ratio $S>5$ is used as detection threshold for the haloes.
Error bars in the upper right panel display standard errors from 10 realisations
for $\tau$CDM at 2, 5, and 10 arcmin (errors for $\gamma_{\rm t}$ offset to the left,
errors for $g_{\rm t}$ offset to the right)
\label{no_noise}}}
\end{figure*}

\begin{figure*}
\resizebox{12cm}{!}{\includegraphics{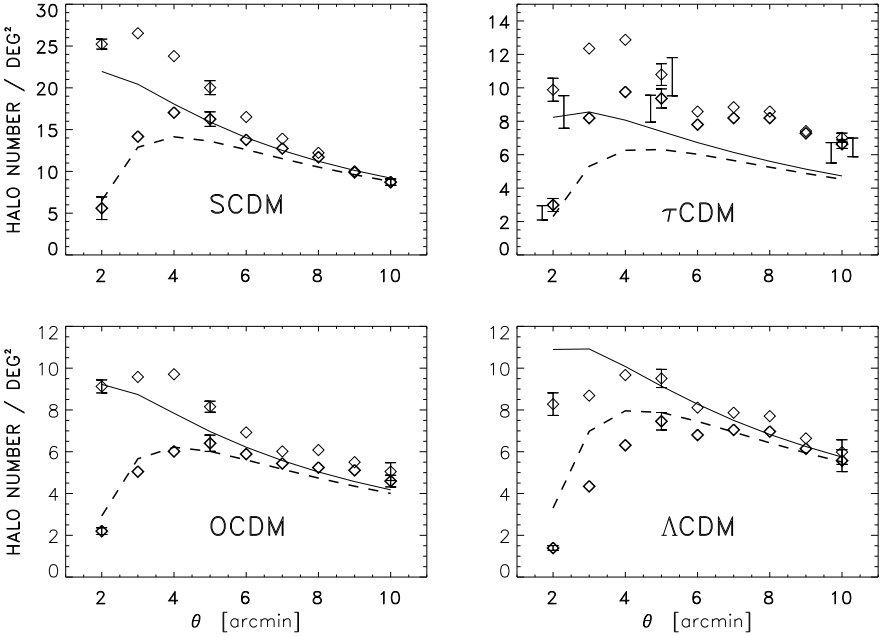}}
\hfill
\parbox[b]{55mm}{
\caption{The same as Fig. \ref{no_noise} but with noise from intrinsic
ellipticities of the sources added. For the theoretical model this is
done by convolving the values from Fig. \ref{no_noise} with a
Gaussian, with the dispersion obtained from the intrinsic ellipticity
distribution (see KS1). Values from the simulations are denoted by
diamonds (thin and thick symbols). Tangential ellipticities now are
obtained according to Eq. (\ref{epsi_t}). A signal-to-noise ratio 
$S>5$ is assumed for the haloes. Error bars centred on the halo abundances
are standard errors for 7 realisations of the ellipticity distribution at 2,
5, and 10 arcmin. Error bars in the upper right panel for $\tau$CDM are from 10
realisations for 2, 5, and 10 arcmin [error for $N(> M_5,\theta)$ offset 
to the right, error for $N(>M_5,>\zeta_{\rm t},\theta)$ offset to the left]. 
\label{with_noise}}}
\end{figure*}

As already indicated in Sect. \ref{map_anal}, high signal-to-noise
peaks of $M_{\rm ap}$ can be identified with dark matter haloes,
rendering the construction of a mass-limited (more correctly:
shear-limited) sample feasible. Analytically, the halo abundances can
be modelled using the Press \& Schechter (1974) prediction for the
mass- and redshift-dependent halo number density, and the universal density
profile of NFW, while in the simulated $M_{\rm ap}$ map 
all connected regions above the corresponding threshold are counted as
haloes. We shall consider haloes with signal-to-noise ratio $S$
larger than 5, i.e., a peak in the $M_{\rm ap}$ map is counted as a halo if
$M_{\rm ap}\ge M_5\equiv 5 \sigma_{\rm c}(\theta)$. 

We consider two differently constructed halo abundances in the following:
The first sample is simply $N(>M_5,\theta)$, the number density of
haloes with an aperture mass larger than $M_5$ for a given
filter size $\theta$. The second sample selects peaks with $M_{\rm
ap}\ge M_5$ within a connected, cross sectional area of $\pi \xi_{\rm t}^2$,
where $\xi_{\rm t}$ is the corresponding cross section radius; the number
density of such peaks is denoted 
$N(>M_5,>\xi_{\rm t},\theta)$. Hence, the size of the peaks in the second
sample exceeds the threshold $\xi_{\rm t}$; these peaks are expected
to be more robust with respect to noise coming, e.g., from the
intrinsic ellipticity distribution and measurement errors. We use a
fixed value of $\xi_{\rm t}=0.6$\thinspace arc minutes. 

In Fig. \ref{no_noise} the number density of the two halo samples as
determined from the simulations without noise are compared to the
results from the analytic calculation in KS1 over
a range of filter scales $2'\le \theta\le 10'$. The four panels in
Fig. \ref{no_noise} refer to the four cosmological models
considered. The error bars for the $\tau$CDM model at 2,5, and 10
arcmin are again obtained from the 10 different realisations centred
on the arithmetic mean of the realisations.

In general, the number counts determined from simulations agree
astonishingly well with the analytical results, considering the simplifying
assumptions entering the latter. The deviations between
simulations and analytical calculation for three of the four cosmologies,
namely SCDM, OCDM, and $\Lambda$CDM, and
especially for the filter scales above 5 arcmin, are less than 10 \%. 
The largest deviation found for these three models is a factor of 2,
for the $\Lambda$CDM model at smallest filter scale.

The only notable exception is the $\tau$CDM model where the deviation
remains above 10 \% even for the largest filter scales ($\theta=10$
arcmin).  This relatively bad agreement has already been noticed for
the rms value of $M_{\rm ap}$ and is probably due to the fact that the
realisation plotted is not characteristic for the mean properties of
that model, as also indicated by the fact that the halo abundance lies
above the mean of all realisations as indicated by the error
bars. 

The good agreement between analytic estimates and numerical results
for the halo number density are surprising, given that (a) 
Press-Schechter theory does not exactly reproduce the spatial number
density of haloes when compared to N-body simulations, and (b) the
universal density profile found by NFW has been obtained by spherical
averaging, and therefore cannot account for the non-axisymmetry of
their projected density. Furthermore, (c) the haloes found from the
simulated $M_{\rm ap}$ are expected to be affected by projection
effects (Reblinsky \& Bartelmann 1999)
which are completely neglected in the analytic
estimates. Despite these effects which one might suspect to yield
significant discrepancies, we find that the analytic estimates are
very accurate. 

We also investigate the halo abundance in an observationally more
realistic situation in Fig. \ref{with_noise}, including the noise from
the intrinsic ellipticity distribution of the background sources. The
plot displays the same quantities as Fig. \ref{no_noise} except for
the fact that the halo abundances of the two different samples have
been determined using the tangential ellipticities in the case of the
simulations. The analytic estimates are obtained as in KS1.  For all
four cosmologies, we determined error bars using 7 different
realisations of the ellipticity distribution of the background sources
(\ref{distrib}).  The error bars from the 10 realisations shown for
$\tau$CDM are slightly sub-Poissonian, as in Fig.  \ref{no_noise}. As
expected from the large value of the kurtosis the error coming from
the intrinsic ellipticity distribution is much smaller than the error
coming from the cosmic variance. On the whole, the number of detected
haloes is increased in all cosmologies because, due to the steepness
of the Press--Schechter mass function for massive objects, there are
more objects just below the threshold than above it. So on average
more objects will be lifted above the threshold by noise than brought
down below it.

\subsection{The tail of $M_{\rm ap}$}

\begin{figure*}[!ht]
\resizebox{12cm}{!}{\includegraphics{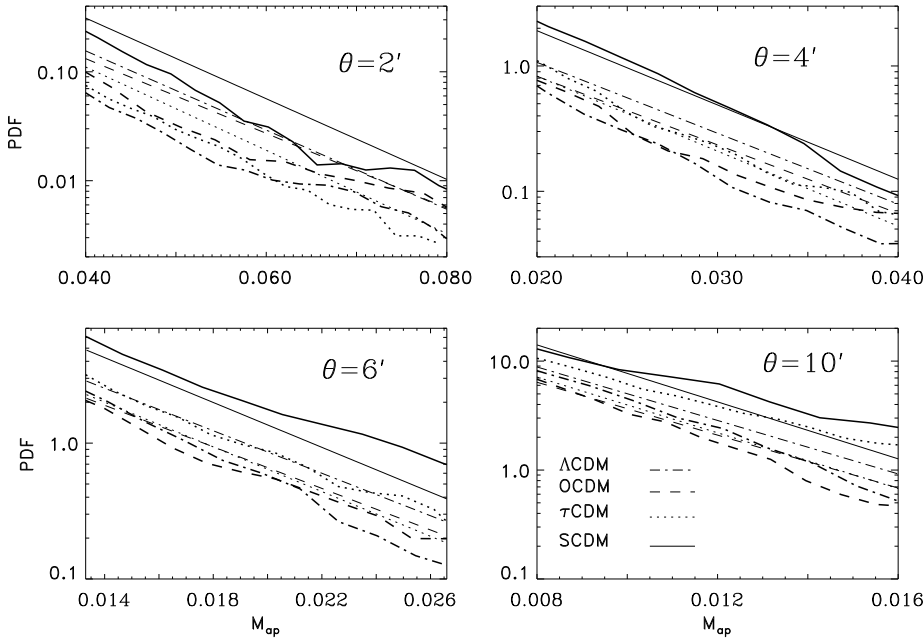}}
\hfill
\parbox[b]{55mm}{
\caption{The tail of the PDF of $M_{\rm ap}$ for the same cosmologies
as indicated in Fig. \ref{histo} for different filter scales.
In each panel we plot the PDF obtained from the analytic estimate in
KS2 with thin lines and
that from simulations with thick lines. The line types specifies the
cosmology: SCDM (solid line), $\tau$CDM (dotted line), OCDM (dashed line) and
$\Lambda$CDM (dashed-dotted line).
The $M_{\rm ap}$-range is $[M_5,2 M_5]$, 
where $M_5=S \times 0.016 / \theta$ with $S>5$ for the parameter
specified in the text.
\label{tail}}}
\end{figure*}

In Fig. \ref{tail} we compare the PDF for $M_{\rm ap}\ge M_5$ as
obtained from analytic calculations (KS2) with
that derived from the simulations without noise.  The PDF is shown for
four filter scales $\theta=2,4,6,10$ arcmin in the range $M_5\le
M_{\rm ap}\le 2 M_5$, for which the analytic results predict a
nearly exponential behaviour. Indeed, the numerical PDF in the
non-Gaussian tail also seems to follow an exponential rather closely,
with a slope very similar to the analytic result.

\begin{figure}[!ht]
\resizebox{\hsize}{!}{\includegraphics{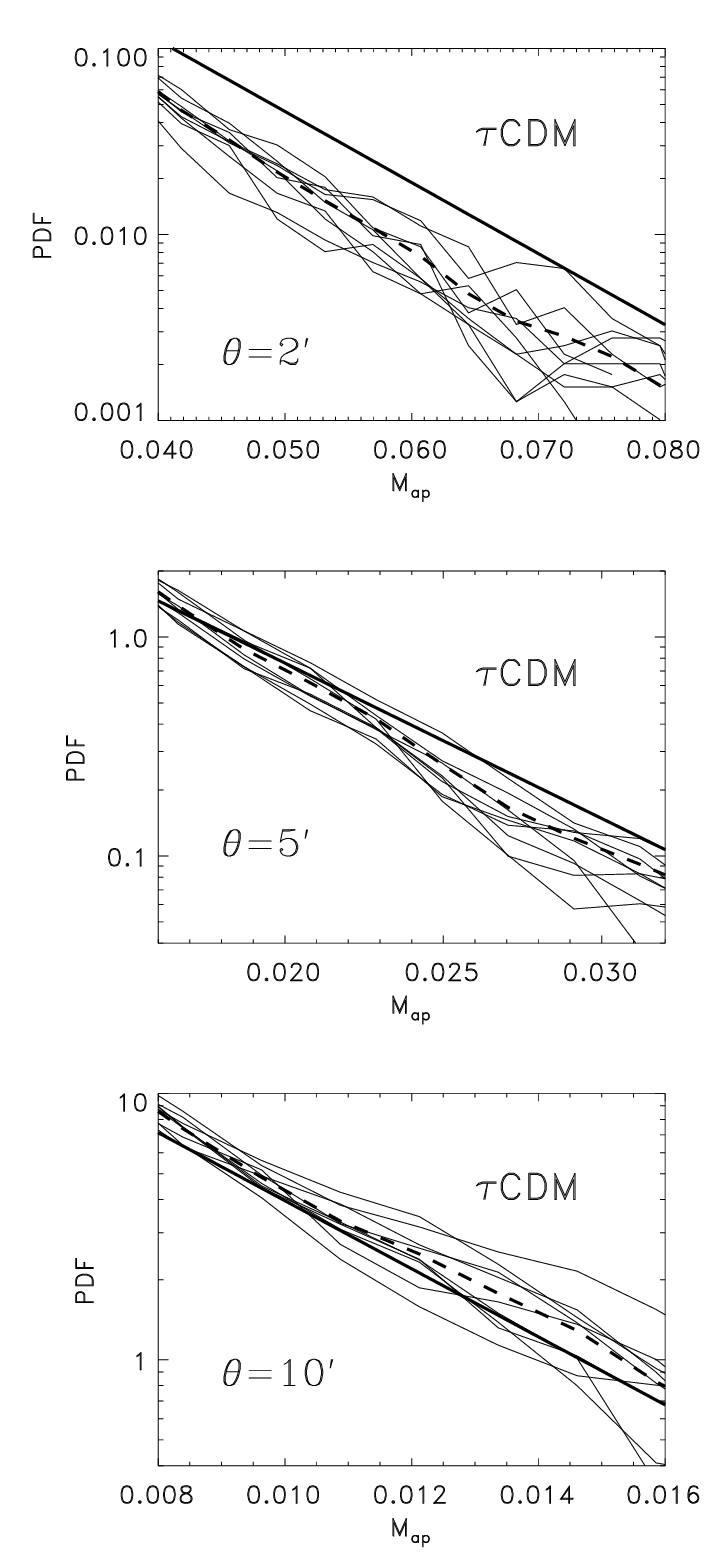}}
\caption{The tail of the PDF of $M_{\rm ap}$ for the 10 realisations
(thin solid lines) of the $\tau$CDM model in comparison to the
analytical result (KS2) for the tail of the PDF of $M_{\rm ap}$ (thick solid
line). In addition, the mean of the 10 realisation (thick dashed line)
is plotted. The different panels are for three filter scales, 2, 5,
and 10 arcmin.
\label{tail1}}
\end{figure}

In order to see how much the PDF varies between different
realisations, we have plotted in Fig. \ref{tail1} the PDF for $M_5\le
M_{\rm ap}\le 2 M_5$ obtained from the 10 realisations in the
$\tau$CDM model, for 3 filter radii, together with their mean and the
corresponding analytic prediction. We find that for the smallest
filter scale $\theta=2'$, all 10 realisations are clearly below the
analytic result, whereas for the larger filters, the realisation
mean of the PDF agrees very well with the analytic prediction.

Remembering that the analytic predictions were made by assuming that
all high values of $M_{\rm ap}$ are coming from regions close to
collapsed haloes, in addition to the assumptions used for estimating
the number density of $M_{\rm ap}$ peaks (Press-Schechter halo
abundance and NFW density profile), this good agreement is somewhat
surprising.

\section{Conclusion}
We used ray-tracing simulations through
N-body-generated cosmic density distributions to study the statistical
properties of the aperture mass $M_{\rm ap}$ as a statistics for
cosmic shear measurements and for finding dark matter haloes from
their shear properties. In particular, we have compared results from
these simulations with the available analytic results and found in
most cases a very good agreement, except for the skewness which is the
least accurate of these predictions. Whereas all other predictions
tested here are based on manifestly non-linear results (like the
Press-Schechter halo abundance and the Peacock \& Dodds power
spectrum), the skewness was estimated analytically by using
second-order Eulerian perturbation theory which, on the scales
considered, is not very accurate. 

Comparing the results from our simulations with analytic studies, we
obtain the following main results: (1) The rms of $M_{\rm ap}$ is
accurately described by analytic results if the fully
non-linear prescription of the power spectrum of density fluctuations
is used. (2) The statistical error of this rms is dominated by
cosmic variance, which in turn depends on the kurtosis of $M_{\rm
ap}$. This kurtosis turns out to be relatively large even on angular
scales of $\sim 10'$, implying the need for many more measurements of
$M_{\rm ap}$ than expected for a Gaussian field, for a given accuracy
of the estimated projected power spectrum. (3) The skewness is only
approximately described by analytic considerations based on
second-order perturbation theory. (4) The predicted abundance of dark
matter haloes detectable at given statistical significance
is very well approximated by the semi-analytic theory which combines
the Press-Schechter number density of haloes with the universal
density profile of Navarro, Frenk \& White. (5) Similarly, the
functional form of the probability distribution of $M_{\rm ap}$ for
values much higher than the rms (i.e., in the non-Gaussian tail) is
found to closely follow an exponential form, of similar slope and
amplitude as predicted by analytic theory which needs to assume that
such high values originate due to collapsed haloes. Thus, on the whole, we
find that the analytical estimates for the statistical properties of
$M_{\rm ap}$ are surprisingly accurate.

We also find that our simulations are not sufficiently large for an
accurate estimate of the higher-order statistical measures, owing to
the finite size of the simulation box in combination with the large
effect of cosmic variance.

As discussed in SvWJK, KS1, KS2, van Waerbeke et
al. (1999) and Bartelmann \& Schneider (1999), the aperture mass is a
useful cosmic shear measure which will eventually allow one to constrain
cosmological parameters, completely independent of any assumption
on the relation between mass and light. For this purpose, the
predictions from cosmology must be known precisely, and our
results here indicate that analytic estimates are relatively
accurate. Unfortunately, we found a large cosmic variance; e.g., in
the estimate of the variance of the rms value of $M_{\rm ap}$, the
kurtosis enters and it decreases only rather slowly with increasing
filter scale. It can be expected that the first successful application
of the aperture mass will be the definition of a sample of haloes
defined in terms of their lensing properties only, with a first
example given by Erben et al.\ (1999). The combination of
cosmic shear information and CMB measurements can be extremely useful,
as shown by Hu \& Tegmark (1999), increasing the precision of the
determination of cosmological parameters substantially over each of
the two individual methods. Their study was based solely on the
dispersion of cosmic shear, i.e., on second-order statistics. It is to
be expected that a similar combination of CMB results with the PDF of
$M_{\rm ap}$ will yield even more precise parameter estimates. A
detailed study of this combination is expected to be very valuable,
but requires a larger grid of cosmological N-body simulations.


\begin{acknowledgements}
We thank M. Bartelmann for his many valuable suggestions and 
a careful reading of the manuscript, as well as an anonymous referee
for his constructive comments.
This work was supported by the ``Sonderforschungsbereich 375-95 f\"ur
Astro-Teilchenphysik" der Deutschen
Forschungsgemeinschaft.
\end{acknowledgements}

\end{document}